\documentclass[pdflatex,sn-mathphys-num]{sn-jnl}

\usepackage{graphicx}%
\usepackage{multirow}%
\usepackage{amsmath,amssymb,amsfonts}%
\usepackage{amsthm}%
\usepackage{mathrsfs}%
\usepackage[title]{appendix}%
\usepackage{xcolor}%
\usepackage{textcomp}%
\usepackage{manyfoot}%
\usepackage{booktabs}%
\usepackage{algorithm}%
\usepackage{algorithmicx}%
\usepackage{algpseudocode}%
\usepackage{listings}%
\usepackage[version=4]{mhchem}

\theoremstyle{thmstyleone}%
\theoremstyle{thmstyletwo}%

\theoremstyle{thmstylethree}%

\makeatletter
\renewcommand{\thetable}{\arabic{table}}
\renewcommand{\fnum@table}{\tablename~\thetable} 
\makeatother

\begin{document}

\title[Article Title]{Structure-Informed Learning of Flat Band 2D Materials}


\author*[1]{\fnm{Xiangwen} \sur{Wang}}\email{xiangwen.wang@manchester.ac.uk}
\equalcont{These authors contributed equally to this work.}

\author[1]{\fnm{Yihao} \sur{Wei}}\email{yihao.wei@postgrad.manchester.ac.uk}
\equalcont{These authors contributed equally to this work.}

\author[1]{\fnm{Anupam} \sur{Bhattacharya}}\email{anupam.bhattacharya@manchester.ac.uk}

\author*[1]{\fnm{Qian} \sur{Yang}}\email{qian.yang@manchester.ac.uk}

\author*[1]{\fnm{Artem} \sur{Mishchenko}}\email{artem.mishchenko@manchester.ac.uk}

\affil[1]{\orgdiv{Department of Physics and Astronomy}, \orgname{The University of Manchester}, \orgaddress{\street{Oxford Road}, \city{Manchester}, \postcode{M13 9PL}, \country{UK}}}


\abstract{
Flat electronic bands enhance electron–electron interactions and give rise to correlated states such as unconventional superconductivity or fractional topological phases. However, most current efforts towards flat-band materials discovery rely on density functional theory (DFT) calculations and manual band structures inspection, restraining their applicability to vast unexplored material spaces. 
While data-driven methods offer a scalable alternative, most existing models either depend on band structure inputs or focus on scalar properties like bandgap, which fail to capture flat-band characteristics.
Here, we report a structure-informed framework for the discovery of previously unrecognized flat-band two-dimensional (2D) materials, which combines a data-driven flatness score capturing both band dispersion and density-of-states characteristics with multi-modal learning from atomic structure inputs. The framework successfully identified multiple flat-band candidates, with DFT validation of kagome-based systems confirming both band flatness and topological character.
Our results show that the flatness score provides a physically meaningful signal for identifying flat bands from atomic geometry. 
The framework uncovers multiple new candidates with topologically nontrivial flat bands from unlabeled data, with consistent model performance across structurally diverse materials.
By eliminating the need for precomputed electronic structures, our method enables large-scale screening of flat-band materials and expands the search space for discovering strongly correlated quantum materials.
}

\keywords{flat bands, 2D material discovery, correlated quantum phases, multi-modal deep learning, structure-based prediction}



\maketitle

\section{Introduction}\label{sec1}

Two-dimensional (2D) materials that host flat electronic bands have become a prime platform for investigating strongly correlated quantum states. When the bandwidth collapses, the electronic kinetic energy is quenched and electron–electron interactions dominate, enabling phenomena such as unconventional superconductivity, ferromagnetism, topological order, and fractional quantum Hall states
\cite{cao2018unconventional,balents2020superconductivity, peri2021fragile,lin2018flatbands,neupert2011fractional, sheng2011fractional,parameswaran2013fractional,liu2014exotic,peotta2015superfluidity}. Representative systems include kagome lattice compounds, Lieb lattice structures, magic-angle twisted bilayer graphene, other moiré superlattices, and other non-moire materials\cite{zhang20252d,mukherjee2015observation, bhattacharya2019flat,bistritzer2011moire,yankowitz2019tuning, tarnopolsky2019origin,he2021moire, chen2020tunable, carr2020electronic,shi2020electronic, zhang2010band, park2023topological}. These examples underscore a central principle: band flatness is a crucial driver of correlated phases. 

Flat bands, however, arise from diverse mechanisms.  Some are trivial, originating from localized atomic orbitals, whereas others result from interference among extended states and may carry non-trivial topology. Distinguishing these cases generally requires detailed analysis of the electronic structure, which limits the scalability of existing discovery workflows. As a result, there is a growing need for general and efficient methods that can detect flat-band features across large databases, regardless of their microscopic origin.

Despite their significance, discovering topological flat-band materials remains difficult because density functional theory (DFT) calculations are computationally intensive. A large variety of previous approaches are based on DFT-calculated bandwidth. For example, Liu et al. \cite{liu2021screening} screened band structure data from 2DMatPedia\cite{zhou20192dmatpedia}; Regnault et al. \cite{regnault2022catalogue} and Duan et al. \cite{duan2024cataloging} applied bandwidth and density of state (DOS) based screening on ICSD materials\cite{vergniory2019complete,vergniory2022all} after high-throughput DFT calculations. These filters all hinge on manually defined selection criteria, such as an arbitrary bandwidth cut-off, which introduces bias. Bhattacharya et al. \cite{bhattacharya2023deep} replaced explicit thresholds with a convolutional neural network that classifies DFT-computed electronic band structure images, but their approach still depends on costly DFT simulations and manually labeled data. However, as all these methods rely on prior DFT calculations, scaling them to unexplored chemical spaces becomes prohibitively expensive. Approaches that forgo DFT are rarer: Neves et al. \cite{neves2024crystal} identified low-dispersion motifs by constructing tight-binding models across the Materials Project database, but the method rests on strong assumptions in the tight-binding parametrization and requires well-defined crystal nets, limiting its generality.

Although data-driven models offer a powerful alternative to traditional electronic structure calculations, most machine learning studies still predict only scalar electronic properties, typically the bandgap or averaged DOS, from crystal structure\cite{zhang2021bandgap, rajan2018machine}. As a result, models built on handcrafted descriptors\cite{knosgaard2022representing}, convolutional neural networks\cite{dong2019bandgap}, graph neural networks, transformer architectures\cite{gong2024graph}, and language models based on textual representations\cite{yeh2025text, lee2025cast} have achieved impressive numerical accuracy. Yet scalar quantities such as bandgap encode only coarse information about electronic dispersion and therefore miss the subtleties needed to pinpoint correlated or topological phases. Consequently, these models remain ill-suited to forward-screening tasks that aim to uncover materials with emergent phenomena such as superconductivity or topology-driven transport.

In this work, we introduce a scalable and interpretable framework for discovering flat bands in 2D materials via crystal structure-informed learning, without requiring any pre-computed electronic structure. By integrating high-throughput inference, sublattice-informed filtering, and embedding-space analysis, our method identifies previously unrecognized flat-band candidates from large-scale materials databases, such as \ce{Nb3TeI7}, \ce{Cu3AsO4}, and \ce{Cu3SbO4} that exhibit fragile topological features near the Fermi level, as confirmed by DFT and band representation analysis. 
Central to our approach is a physics-motivated flatness score, derived from electronic band dispersion and DoS features, which serves as an interpretable supervision signal grounded in materials physics. This score is predicted directly from atomic structures using a multi-modal deep learning model trained on aligned graph and text encoders. Scalability is achieved by applying the model to over ten thousand candidate structures without requiring explicit band structure inputs, enabling efficient screening across large chemical spaces.
This study provides a physically grounded, data-efficient pathway for discovering flat-band materials directly from atomic structure, offering a scalable route toward data-driven discovery of correlated quantum materials.

\section{Results}\label{sec:results}
\begin{figure*}
\centering
\includegraphics[width=0.95\textwidth]{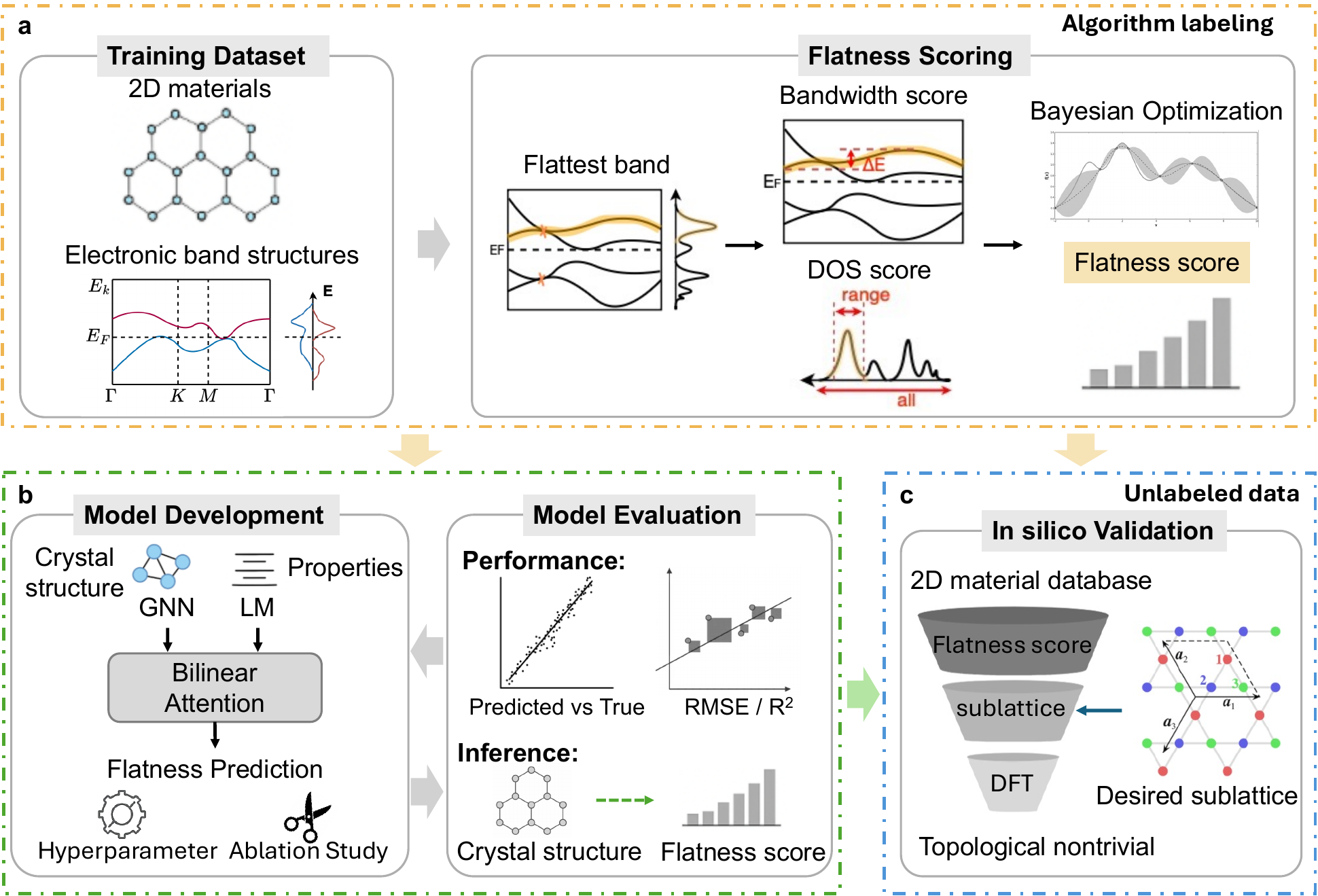}
\caption{\textbf{Framework for automatic labeling and identification of flat-band phases in 2D materials using structure-informed algorithms.}
\textbf{a} Dataset labeling and flatness scoring. A composite flatness score combining bandwidth and DOS characteristics is used to automatically label training data based on electronic band structures.
\textbf{b} Multi-modal model development. A multi-modal model trained on structural and textual features predicts flatness scores from crystal structure alone, with performance validated on labeled data.
\textbf{c} Application to unlabeled materials. The model is applied to unlabeled materials, and high-scoring candidates are filtered by kagome-like sublattice motifs and validated via DFT calculations.
}
\label{fig:overview}
\end{figure*}

Fig.~\ref{fig:overview} outlines the workflow for large-scale discovery of flat-band 2D materials.
First, materials with known electronic structures are labeled via a flatness score combining band dispersion and DoS characteristics, capturing essential spectral signatures of flat-band behavior (Fig.~\ref{fig:overview}a). This algorithmic labeling provides a continuous and physically grounded supervision signal for training a multi-modal deep learning model on atomic structural inputs, combining geometric and contextual representations (Fig.~\ref{fig:overview}b). The trained model is then applied to a broad set of unlabeled 2D materials to rank candidates by predicted flatness. High-scoring structures are further screened through sublattice motif analysis and validated by DFT calculations, confirming the presence of topologically nontrivial flat bands in multiple cases (Fig.~\ref{fig:overview}c).
By learning structure–property relationships grounded in physically defined flatness criteria, the framework enables large-scale screening of candidate materials without electronic structure inputs, while preserving meaningful connections to underlying band topology.

\subsection{High-throughput Scoring of Band Flatness in 2D Materials}
To systematically identify flat-band features in 2D materials, we constructed a dual-metric framework to automatically generate flatness scores based on electronic band dispersion and DOS characteristics, without manual annotation. We applied this procedure to the 2DMatPedia database\cite{zhou20192dmatpedia}, which provides computed electronic structures for around 5,100 2D materials.

\begin{figure*}
\centering
\includegraphics[width=1\textwidth]{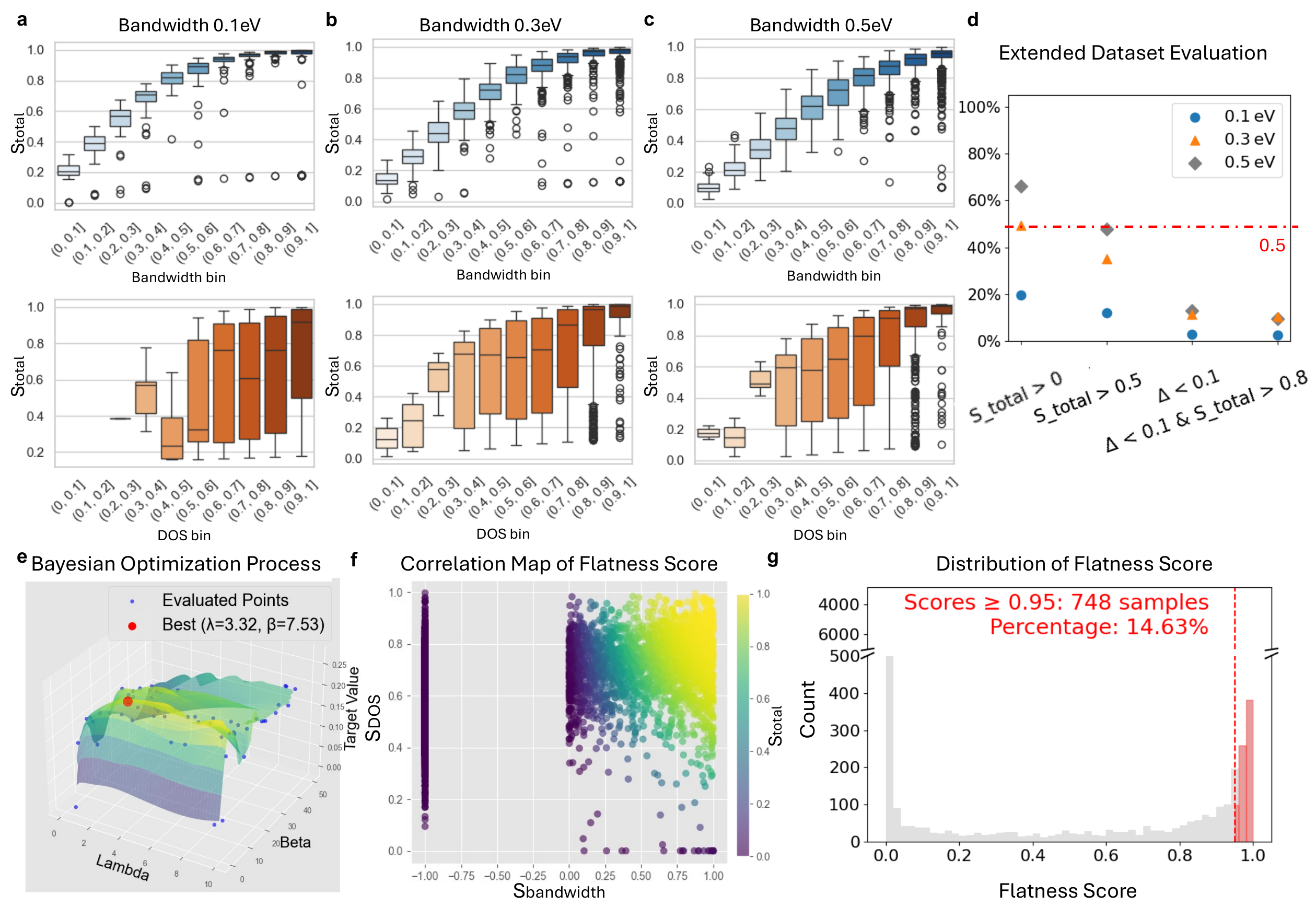}
\caption{\textbf{Flatness Scoring Optimization and Analysis of Electronic Structure Data.}
\textbf{a-c}  Box plots of different bandwidth windows (0.1 eV, 0.3 eV, 0.5 eV) showing the distribution of $S_{\text{total}}$ for samples with $S_{\text{total}} > 0$, grouped by intervals of $S_{\text{bandwidth}}$ (top panel) and $S_{\text{DOS}}$ (bottom panel); color intensity reflects the number of samples in each bin to indicate data balance across intervals.
\textbf{d}  Evaluation of dataset-level quality metrics across three different bandwidth windows (0.1 eV, 0.3 eV, 0.5 eV). $\Delta$ refers to the absolute difference between the two sub-scores, defined as $|S_{\text{bandwidth}} - S_{\text{DOS}}|$.
\textbf{e} Bayesian optimization process with evaluated points (blue) and best parameters (red). 
\textbf{f} Correlation Map of Flatness Score, $S_{\text{bandwidth}}$ and $S_{\text{DOS}}$, colored by $S_{\text{total}}$. 
\textbf{g} Distribution of $S_{\text{total}}$ scores across the whole dataset, with a threshold of 0.95 used to show the top high-quality candidates.} 
\label{fig:BO}
\end{figure*} 

For each material, we first detect band crossings, then divide the band structure -- between consecutive high-symmetry points -- into continuous, non-crossing segments. These segments are then reconnected at crossing points to form multiple end-to-end combinations, from which the one with the narrowest energy span is chosen as the representative band the material.

We define a momentum-space flatness score, $S_{\text{bandwidth}}$, to quantify the dispersion of the material. Specifically, we first compute the energy span of the previously identified representative band, and then map the raw value to the interval $[0,1]$ using a cosine transformation that penalizes broader bands: a perfectly flat segment approaches one, while highly dispersive bands approach zero. Building on this definition, we introduce a tunable threshold $\omega_{\max}$ to specify the maximum energy span still considered “flat”; the energy span wider than this limit automatically receive zero scores.  We intentionally avoid an overly strict cut-off (for example, 0.05 eV), which would shrink the training set and reduce the problem to a binary classification. By retaining a continuous, regression-style score, the model preserves subtle gradations of flatness and enables fine-grained prioritization of candidate materials.

To complement this dispersion-based metric, we define a density-of-states measure, $S_{\text{DOS}}$, which recognizes the pronounced DOS peaks typical of flat-band systems\cite{heikkila2011flat,li2018realization}. For representative band of every material, we center a window of width $\omega_{\max}$ on the band midpoint, calculate the mean DOS inside that window, and compare it with the DOS of the fixed reference range of [-5 eV, 5 eV] \cite{regnault2022catalogue} . The contrast is then normalized to $[0,1]$, where higher values indicate sharper peaks. A saturation function mitigates the influence of low background DOS, stabilizing the score. 

The overall flatness metric, $S_{\text{total}}$, combines $S_{\text{bandwidth}}$ and $S_{\text{DOS}}$ under a "both-must-be-high" principle: if $S_{\text{bandwidth}}$ falls below $\omega_{\max}$, $S_{\text{total}}$ is set to zero regardless of the DOS peak. Rather than manually defining the combination weights, we learn them via Bayesian Optimization (BO), letting the data tune the parameters to best reflect the underlying physics.

Although our flatness metrics are continuous, their effectiveness depends on the bandwidth threshold $\omega_{\max}$. We evaluated three values, 0.1, 0.3, and 0.5 eV, to examine how different thresholds influence both the score distribution and sample balance. As shown in Fig.~\ref{fig:BO}a–c, when $\omega_{\max}$ is set to 0.3 or 0.5 eV, $S_{\text{total}}$ increases more smoothly across binned $S_{\text{bandwidth}}$ and $S_{\text{DOS}}$ values, with more balanced sample distributions as indicated by the color intensity. In contrast, the 0.1 eV setting causes the scores to saturate at low values, resulting in sparsely populated high-score regions and limited usable data coverage. Fig.~\ref{fig:BO}d summarizes the proportion of structures meeting increasingly strict quality criteria across the full dataset. These include basic validity ($S_{\text{total}} > 0$), high flatness confidence ($S_{\text{total}} > 0.5$), structural balance (measured by $\Delta = |S_{\text{bandwidth}} - S_{\text{DOS}}| < 0.1$), and a combined constraint that selects only those structures with both high scores and balanced contributions ($\Delta < 0.1$ and $S_{\text{total}} > 0.8$). While 0.5 eV yields the highest number of candidates, 0.3 eV provides comparable coverage and imposes a stricter physical constraint on band dispersion, enhancing selectivity. In contrast, the 0.1 eV threshold is too stringent, resulting in limited usable data, eliminating most samples under any realistic constraint. 

Fig.~\ref{fig:BO}e shows the optimized parameter space obtained using a bandwidth threshold of 0.3 eV, where $S_{\text{total}}$ peaks in regions with simultaneously high $S_{\text{bandwidth}}$ and $S_{\text{DOS}}$, while penalizing imbalance. This results in a nontrivial scoring surface that preserves resolution across the dataset and highlights high-quality flat-band candidates. Fig.~\ref{fig:BO}f visualizes the joint distribution of $(S_{\text{bandwidth}}, S_{\text{DOS}})$, color-coded by $S_{\text{total}}$. High-scoring samples cluster in the top-right, confirming that the learned combination captures the desired trade-off. Entries that exceed the bandwidth threshold ($S_{\text{bandwidth}} = -1$) appear as a vertical strip with $S_{\text{total}} = 0$. The overall score distribution in Fig.~\ref{fig:BO}g shows that, while a substantial portion of materials are filtered out, the remaining entries span a broad range. Notably, 748 materials (14.63\%) achieve scores above 0.95 and are shortlisted for further theoretical and structural investigation. Results for other thresholds (0.1 and 0.5 eV) are provided in Supplementary Section A.

\subsection{Deep Learning Model for Flatness Score Prediction}

\begin{figure*}
\centering
\includegraphics[width=0.98\textwidth]{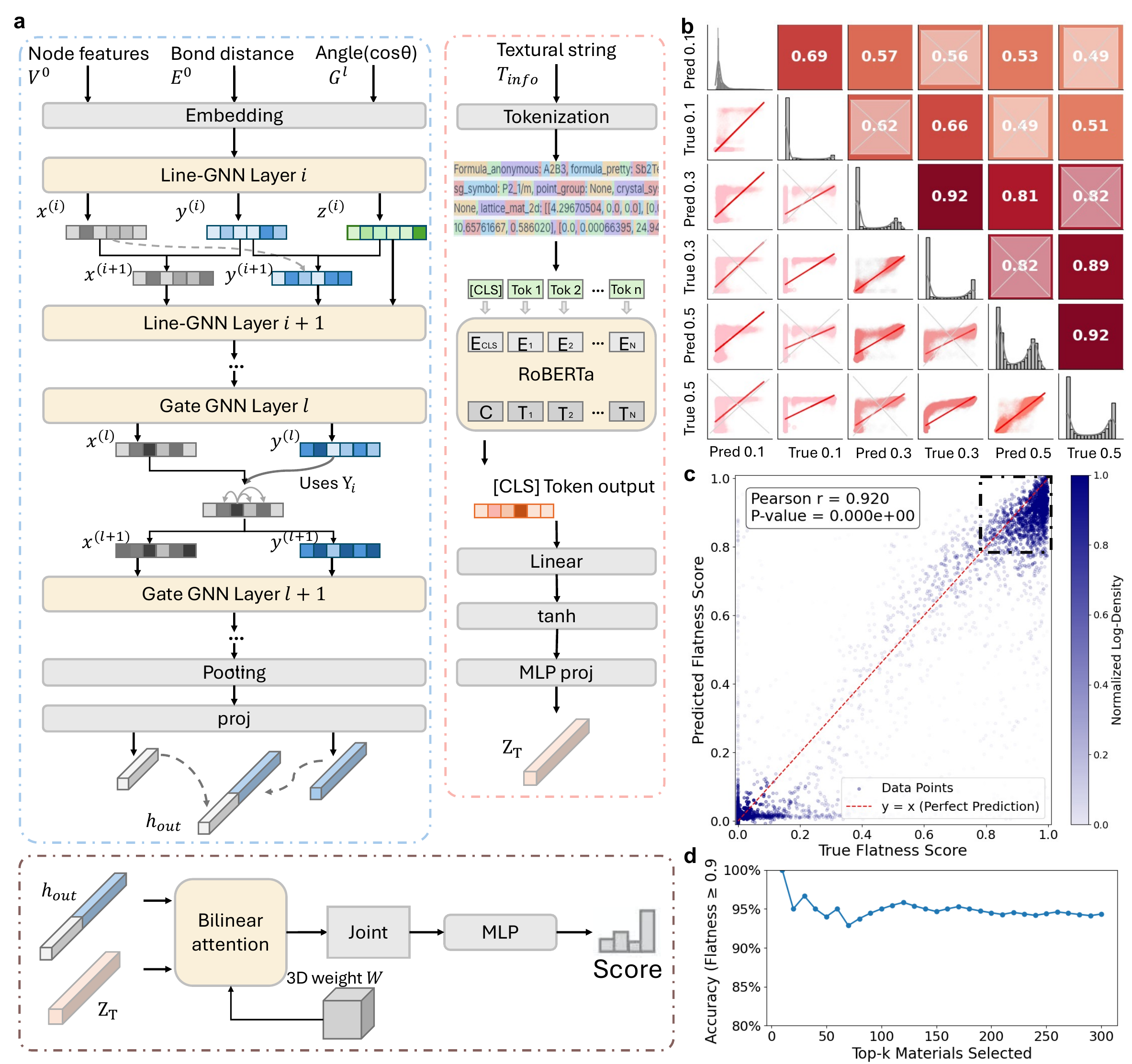}
\caption{\textbf{Multi-modal architecture for predicting flatness scores and assessing model performance.}
    \textbf{a} Model overview from crystal structure to flatness score prediction. Atomic structures are encoded using a gated line graph neural network, while textual material descriptions are processed via a RoBERTa encoder. The two modalities are fused through bilinear attention to produce a continuous flatness score.
    \textbf{b} Pairwise relationships between predicted and true values across experiments with bandwidth thresholds $\omega_{\max}$ of 0.1 eV, 0.3 eV, and 0.5 eV. Diagonal plots display the distribution of each variable (histogram + KDE). Lower triangle plots show scatter plots with regression lines (red); point transparency reflects density. Upper triangle plots present Pearson correlation coefficients, with background color intensity indicating correlation strength (darker for stronger).
    Light gray 'X' markers denote non-comparable pairs.
    \textbf{c} Model performance: predicted vs. true flatness scores. 
    \textbf{d} Top-k selection accuracy for flat-band materials (flatness score $\geq 0.9$). The y-axis denotes the proportion of true flat-band materials among the top-k ranked predictions.
} 
\label{fig:dl}
\end{figure*}

\subsubsection{Multi-modal Deep Learning Framework}
To directly predict the flatness score from raw material inputs, we constructed a multi-modal deep learning framework that integrates structural and textual information of 2D material properties. 

The structural encoder is based on the ALIGNN framework\cite{choudhary2021atomistic}, which augments graph neural networks with line-graph-based geometric priors, capturing bond lengths and angular relations through edge-gated message passing. As illustrated in Fig.~ref{fig:dl}a (left), the model processes atomistic features across stacked Line-GNN and Gate-GNN layers to extract hierarchical descriptors of 2D crystal geometry. In parallel, chemical formulas and structured descriptors are embedded using a transformer-based encoder (Fig.~\ref{fig:dl}a, right), where a pre-trained RoBERTa model\cite{liu2019roberta} generates contextualized token-level embeddings. These are projected into a fixed-dimensional latent vector. The two modalities are fused via a bilinear attention module, which learns cross-modal interactions between structural and textual embeddings. The final joint representation is used to predict the flatness score. This end-to-end model allows prediction from unprocessed material inputs, eliminating manual feature engineering while capturing both local structural features and semantic material context relevant to flat-band behavior.

\subsubsection{Model Performance on Labeled Dataset}
After training on the labeled dataset, we systematically evaluated the predictive performance of our multi-modal model. Fig.~\ref{fig:dl}b presents pairwise comparisons between predicted and true flatness scores across the three bandwidth thresholds (0.1 eV, 0.3 eV, and 0.5 eV), showing both within-threshold accuracy and cross-threshold consistency. 
Under each individual bandwidth setting, the Pearson correlation coefficients between predicted and true values were 0.66 for 0.1 eV, and 0.92 for both 0.3 eV and 0.5 eV, indicating comparable predictive accuracy between the latter two, and a substantial improvement over the stricter 0.1 eV threshold. Cross-setting comparisons also show strong correlations, particularly between the 0.3 eV and 0.5 eV conditions, where the true-value correlation reaches 0.89 and the predicted-value correlation is 0.81. These results suggest that the model outputs are robust to bandwidth variation, with especially high consistency between the 0.3 eV and 0.5 eV settings. Taken together, these findings indicate that a threshold of 0.3 eV achieves the same predictive fidelity as 0.5 eV, while avoiding the excessive permissiveness associated with broader thresholds. It thus offers a more desirable trade-off between precision and generalizability. All subsequent analyses are therefore based on models trained under the $\omega_{\max} = 0.3$ eV bandwidth setting.

Fig.~\ref{fig:dl}c further examines the predictive accuracy under the 0.3 eV setting. The predicted scores exhibit strong agreement with ground truth, with a Pearson correlation of 0.92 and a p-value effectively equal to zero. Predictions closely follow the identity line, indicating that the model effectively learns the distribution of flatness scores and captures subtle variations across diverse 2D materials. Notably, the model consistently identifies highly flat-band systems, as reflected by the clustering of high-score predictions in the upper right region. To assess the model's effectiveness in practical screening, we examined the proportion of true flat-band materials (with true scores higher than 0.9) among the top-k candidates. As shown in Fig.~\ref{fig:dl}d, the accuracy remains close to 95\% within the top 150, far above the random baseline of 50\%. This highlights the model's ability to reliably prioritize high-quality candidates for flat-band materials discovery.

Beyond prediction accuracy, we also examined how the model internally organizes structural information. To probe the learned structural representations, we projected the high-dimensional embeddings from the graph encoder using uniform manifold approximation and projection (UMAP)\cite{mcinnes2020umapuniformmanifoldapproximation}. Fig.~\ref{fig:cluster}a shows that materials with similar flatness scores form coherent clusters in the embedded space. Low-flatness and high-flatness systems occupy well-separated regions, indicating that the encoder organizes materials according to latent structural factors correlated with flat-band behavior, providing an interpretable blueprint for subsequent materials exploration.

\subsection{Screening Flat-Band Materials and DFT Validation}

\begin{figure*}
\centering
\includegraphics[width=0.98\textwidth]{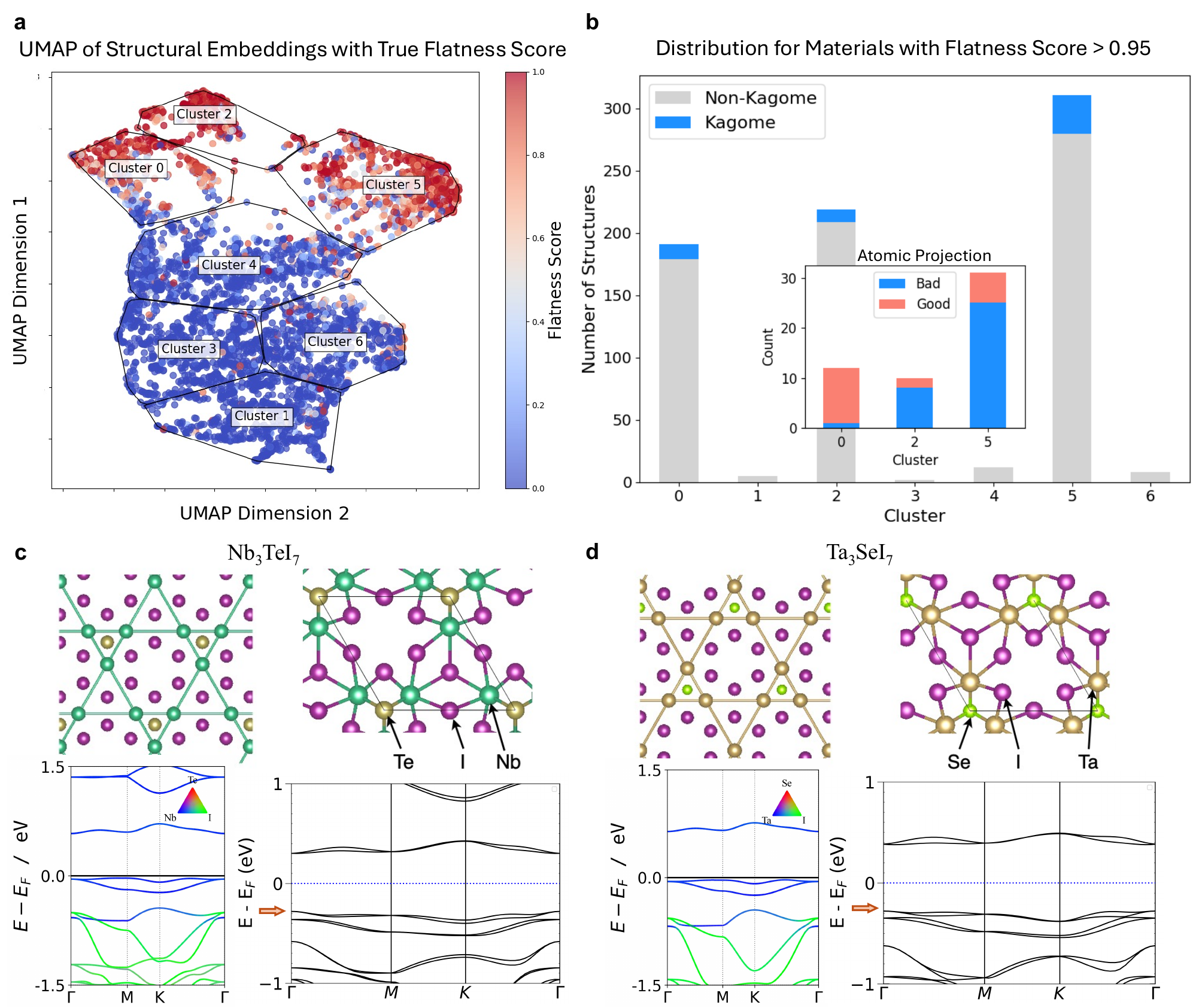}
\caption{\textbf{Screening flat-band materials from labeled database and first-principles validation.}
\textbf{a} UMAP visualization and clustering results of the structure embeddings of 2D materials from 2dmatpedia dataset generated by the deep learning model. 
\textbf{b} The cluster-wise distribution and average flatness scores. 
\textbf{c} DFT-calculated band structure and atomic configuration of \ce{Nb3TeI7}. The breathing kagome sublattice formed by Nb atoms is shown in green (top left), with the unit cell on the right. Element-projected bands without SOC (bottom left) indicate that the flat band near the Fermi level is mainly supported by Nb orbitals. The calculated band structure involving SOC effect is shown in the bottom right panel. The flat bands highlighted by a red arrow are separated from nearby bands under SOC and exhibit fragile topology.
\textbf{d} DFT-calculated band structure and atomic configuration of \ce{Ta3SeI7}. The flat bands denoted by the red arrow are supported by Ta atoms in the breathing kagome sublattice and also isolated with spin–orbit coupling, of which the non-trivial topology is demonstrated.
} 
\label{fig:cluster}
\end{figure*}

To prioritize candidates for analysis, we first selected all materials with predicted flatness scores above 0.95 and mapping their locations in the UMAP projection (Fig.~\ref{fig:cluster}b). These high-flatness materials formed distinct clusters, within which we focused on structures containing kagome motifs due to their known propensity for hosting topological flat bands. We then examined the orbital-projected band structures of these candidates and retained only those where the near-Fermi flat band is predominantly derived from atoms in the kagome sublattice (see inset of Fig.\ref{fig:cluster}b). Materials failing this atomic projection criterion were discarded. A complete catalog of high flatness score materials, including cluster assignment, kagome-sublattice annotation, and orbital projections, is available on our Github repository. To simplify topological characterization, we excluded materials labeled as "magnetic" in the 2DMatPedia database, avoiding the additional complexity introduced by magnetic ordering. The screening distilled our pool to 15 high-confidence flat-band candidates, primarily located in clusters 0, 2 and 5, which we then subjected to for further theoretical investigation.

The top ranked materials were investigated by DFT calculations to verify the presence of topological flat bands. One representative case is \ce{Nb3TeI7} (2dmatpedia ID 2dm-3841). Its breathing kagome Nb sublattice hosts almost dispersionless band ($\omega \approx 0.05$ eV) near the Fermi level (Fig.~\ref{fig:cluster}c). Inclusion of spin-orbit coupling (SOC) separates the flat band from its neighbor and lifts the spin degeneracy everywhere except at $\Gamma$ point. Topological quantum chemistry analysis was performed using elementary band representations (EBR)\cite{petralanda2024two, bradlyn2017topological}. \ce{Nb3TeI7} belongs to layer group 69 (P3m1). The spin-orbit-split doublet transforms as $\Gamma_4 \oplus \Gamma_5$, $K_4 \oplus K_5$, and $M_3 \oplus M_4$ at high symmetry points. This set of irreducible representations cannot be expressed as a positive sum of the layer-group EBRs; it can only be written as a difference between them, a hallmark of fragile topology, akin to the flat bands in magic-angle twisted bilayer graphene \cite{cao2018unconventional, peri2021fragile, song2019all}. The topological flat bands are supported by \ce{Nb} atoms forming a breathing kagome sublattice structure, as a result of destructive interference of wavefunctions associated with this geometry. Notably, a chemically similar system, \ce{Nb3Cl8}, has been experimentally demonstrated to have topological flat bands\cite{sun2022observation}. 

Several additional high-flatness candidates display the same fragile kagome-derived topology. \ce{Ta3SeI7} (2dm-5470), with structure closely resembling that of \ce{Nb3TeI7} (Fig.~\ref{fig:cluster}d), hosts an SOC-isolated flat band with the same band representation characteristics. Two further layer-group-69 compounds, \ce{Ta3SBr7} (2dm-5348) and \ce{Ta3TeI7} (2dm-5496), likewise possess Ta-dominated flat bands pinned within a few tens of meV of the Fermi level, generated by their breathing kagome Ta sublattices. Complete band structures and atomic configurations for \ce{Ta3SBr7} and \ce{Ta3TeI7} are provided in Supplementary Section C.

\subsection{Application on Unlabeled Dataset and Validation}

\subsubsection{Flatness Score Prediction and Structural Clustering}
To explore the utility of our flatness prediction model, we deployed it on an unlabeled dataset comprising 2D materials from the C2DB database\cite{gjerding2021recent}. For each structure, the model predicted a flatness score in a single forward pass using only the crystal geometry as input. No electronic structure information was supplied, making this a fully out-of-distribution test of the learned structure-property mapping. 

\begin{figure*}[h!]
\centering
\includegraphics[width=0.97\textwidth]{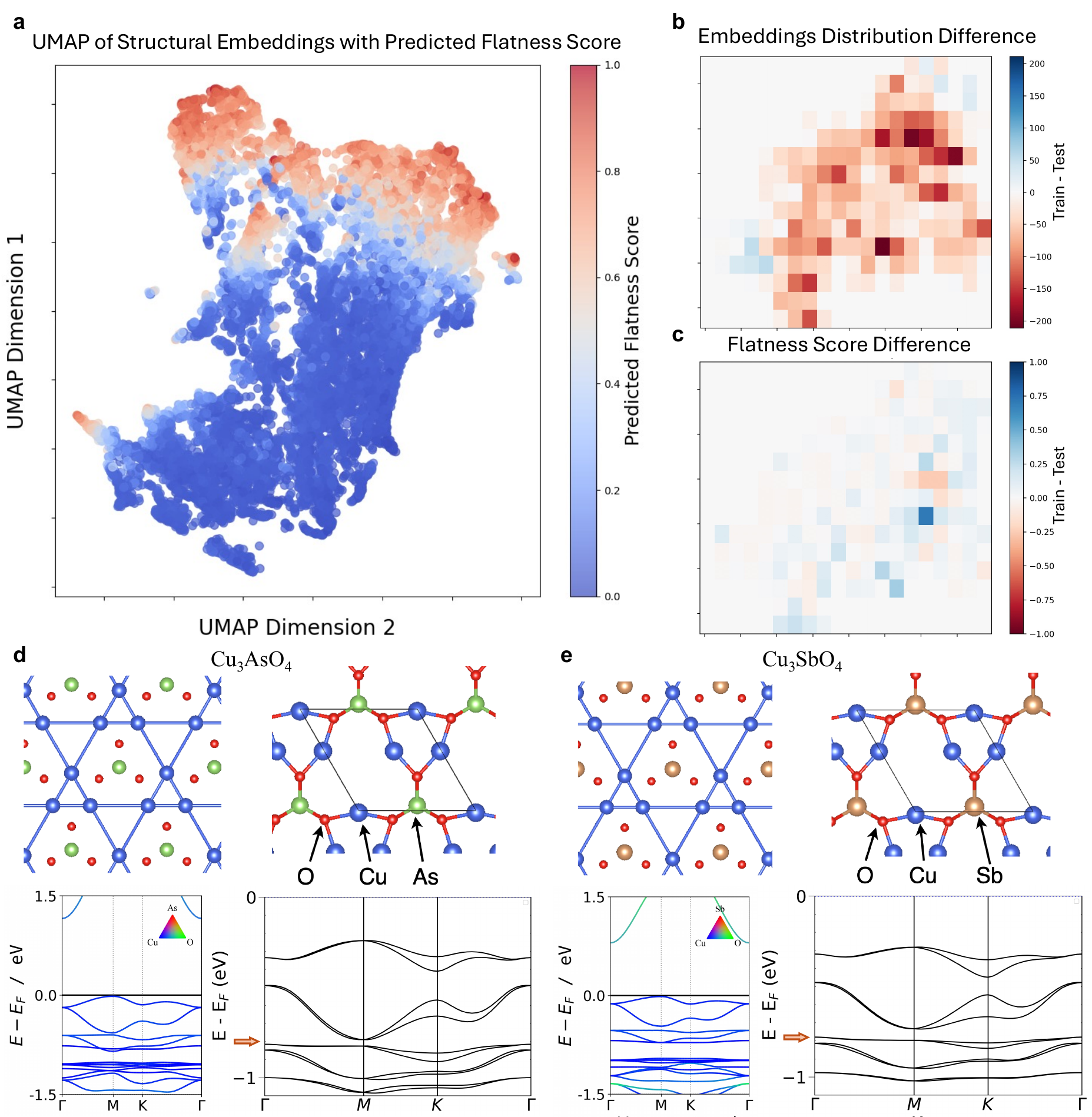}
\caption{\textbf{Analysis of unlabeled dataset by the deep learning model and first-principles validation.}
\textbf{a} UMAP projection of structural embeddings from the inference dataset, colored by predicted flatness score.
\textbf{b} Comparison between training and inference datasets in structural embedding space for  count difference (Train - Inference). 
\textbf{c} Comparison between training and inference datasets for flatness score difference (Train - Inference).
\textbf{d} Crystal structures and electronic band structures of \ce{Cu3AsO4}. Breathing kagome sublattice is shown in the top left panel connecting Cu atoms. To the right is the primitive unit cell. The bottom left panel shows the elemental projected band structure without SOC. Electronic spectrum with SOC is presented in the bottom right panel. \textbf{e} Crystal structure and electronic band structure of \ce{Cu3SbO4} similar to that of \ce{Cu3AsO4} in \textbf{d}.
}
\label{fig:c2dbdl}
\end{figure*}

Fig.~\ref{fig:c2dbdl}a shows the UMAP projection of structure embeddings of materials from the unlabeled dataset (C2DB), colored by the predicted flatness scores. The distinct color gradient demonstrates that the learned embeddings effectively capture the geometric features relevant to flat-band behavior. High-scoring materials exhibit clear clustering, highlighting the model's strong generalization to previously unseen structural domains. A complete catalog, including flatness score, cluster assignment, and kagome-lattice flag for every top-ranked material, is available on our GitHub page. 

To further assess the model’s generalization capability, we compare the distribution of structural embeddings obtained during training (Fig.~\ref{fig:cluster}a) and inference (Fig.~\ref{fig:c2dbdl}a), and examined corresponding deviations in the flatness scores across the embedding space. This analysis evaluates whether shifts in the embedding distribution occur when the model is applied to unseen data, and whether such shifts affect predictive performance. It also facilitates a comparison between the training and inference datasets, which are otherwise difficult to align due to differences in labeling and structural diversity.

The embeddings produced by the graph encoder were projected into two dimensions using UMAP, followed by spatial discretization into grid cells. For each cell, we computed the difference in sample density between training and inference sets (Fig.~\ref{fig:c2dbdl}b), as well as the difference in the average predicted flatness score (Fig.~\ref{fig:c2dbdl}c).
A shift in embedding distribution is evident, with several peripheral and central regions showing a higher density of inference samples relative to the training set. These regions correspond to structural motifs that were underrepresented during training and, in some cases, are associated with low predicted flatness scores.
Despite this distributional shift, the predicted flatness scores remain largely consistent across the embedding space. Only a few localized regions exhibit noticeable deviations, indicating that the model retains its predictive stability even in domains not encountered during training.

\subsubsection{DFT Validation of Predicted High-Scoring Materials}
To validate the model's predictions, we performed sublattice-guided screening followed by DFT calculations on a subset of high-scoring, non-magnetic materials from the C2DB set. Specifically, 19 candidates exhibiting kagome-like substructures were selected and their electronic structures were calculated without spin-orbit coupling. The band structures are plotted to confirm band flatness, and atomic projection of the corresponding flat bands is analyzed to exclude flat bands unassociated with atoms in kagome sublattices. This process led to the discovery of several previously unreported topological flat-band candidates.

Among them, \ce{Cu3AsO4} and \ce{Cu3SbO4} emerge as two structurally similar systems with nontrivial flat-band features. Both materials belong to layer group 69 (P3m1). The irreducible representations at high symmetry points $\Gamma$, $K$ and $M$ in the momentum space are calculated and compared with irreps of EBRs. As shown in Fig. \ref{fig:c2dbdl}d–e, the third highest valence band lies close to the Fermi level and is spectrally isolated due to spin-orbit coupling. The decomposition of band representations is $\Gamma_4 \oplus \Gamma_5$, $K_5 \oplus K_6$, and $M_3 \oplus M_4$, which do not correspond to any linear combination of irreps of EBRs. Through summation of representations of this topological band and the two bands below which have trivial topologies, a linear combination of irreps of EBRs can be built, suggesting the fragility of the non-trivial topology. These features mirror the topological flat bands found in kagome systems, reinforcing the method’s capability to uncover novel correlated quantum materials.

\section{Discussion}\label{sec12}




Our structure-informed framework represents a methodological advance in quantum materials screening by combining interpretable supervision with structure-property representation learning. The physics-motivated flatness score provides a transparent and physically grounded training signal, while the multi-modal encoder captures latent structure–property relationships directly from atomic structure inputs. In the resulting latent space, UMAP reveals a clear separation between low- and high-flatness materials, with kagome-like motifs clustering in specific regions, consistent with their known role in flat-band formation. Importantly, these structure–score relationships persist across both the training set and unseen materials, suggesting that the model captures transferable geometric features rather than dataset-specific correlations. Although the latent distribution shifts between training and inference sets, predicted flatness scores remain remarkably stable, reinforcing the model’s ability to generalize across chemically and structurally diverse materials. This robustness enables scalable flat-band discovery across vast materials spaces, without resorting to computationally intensive electronic structure calculations.

The framework uncovered several previously unreported 2D materials hosting isolated flat bands with fragile topological character, unified by kagome-like sublattice motifs and symmetry features associated with nontrivial band topology. These materials share breathing kagome sublattices and belong to layer group 69 (P3m1), emphasizing the geometric motifs underpinning flat-band formation. In particular, \ce{Nb3TeI7}, \ce{Ta3SeI7}, \ce{Ta3SBr7} and \ce{Ta3TeI7} exhibit spin–orbit-induced band isolation for the first degenerate bands below their Fermi levels and representations of the symmetry group inconsistent with EBR, indicating non-trivial topology, while \ce{Cu3AsO4} and \ce{Cu3SbO4}, both from unlabeled datasets, validate the model's ability to generalize beyond the training distribution. These results highlight the effectiveness of the flatness score and structure-based modeling in revealing hidden structure–property relations linked to flat-band behavior.

\section{Conclusions}
We have presented a structure-informed framework that enables high-throughput discovery of flat-band 2D materials without requiring electronic structure inputs.
By integrating a physics-based flatness score with a multi-modal deep learning model, we demonstrated its ability to identify previously unreported topologically nontrivial flat-band candidates from large-scale unlabeled databases. 

Although here we focused on flat bands, the approach is readily extensible. By redefining the scoring function to target, for example, exchange splittings or symmetry indicators, the same framework could accelerate the search for 2D magnets, spin-liquid platforms, or other quantum phases. Its generalizability also supports application to 3D layered compounds and van der Waals heterostructures using only atomic inputs. While the current sublattice filter focuses on kagome motifs, future extensions could broaden motif coverage and incorporate symmetry-aware architectures to capture a wider range of flat-band geometries and related emergent phenomena.

\section{Methods}\label{sec11}

\subsection{Flatness Scoring} \label{m1}
To systematically identify flat bands from the band structures of reported 2D materials, we develop a physically motivated flatness evaluation method. This approach quantifies the energy dispersion of bands across the Brillouin zone, augmented by the scores of DOS peaks to capture the concentration of electronic states and their proximity to the Fermi level, providing a foundation for subsequent machine learning predictions.

\subsubsection{Band Selection and Fermi Level Relevance}
Flat bands are physically most significant when located near the Fermi level, as they directly influence the low-energy physics of the material. Accordingly, we select a subset of bands from the material's band structure that are closest in energy to the Fermi level. 
Specifically, for a given 2D material, we compute the average energy of each band and rank them based on their energy difference from the Fermi level, choosing the $n$ bands with the smallest differences (or all bands if fewer than six are available). For spin-polarized materials, spin-up and spin-down channels are evaluated separately to capture spin-dependent features relevant to magnetic or strongly spin–orbit-coupled systems.

To capture their dispersion characteristics, we examine these bands along continuous $k$-point paths in the Brillouin zone, constructed by connecting high-symmetry points. Each selected band is further segmented into continuous, non-crossing segments by identifying crossings with adjacent bands at each $k$-point. These segments are then reconnected at crossing points to form multiple full dispersion curves along high-symmetry paths. The curve with the narrowest energy span is then identified as the representative band for flatness evaluation of the material. 

\subsubsection{Definition of Flatness Score}
We define a composite score $S_{\text{total}}$ that integrates $S_{\text{bandwidth}}$ and $S_{\text{DOS}}$, emphasizing minimal dispersion and high state density near the Fermi level.

The bandwidth component, $S_{\text{bandwidth}}$, directly quantifies the flatness of the representative band as follows:
\begin{equation}
S_{\text{bandwidth}} = 
    \begin{cases} 
        -1, & \text{if } \Delta E > \omega_{\text{max}} \\
        \frac{1}{2} [\cos(\pi \cdot \Delta E / \omega_{\text{max}}) + 1], & \text{otherwise} 
    \end{cases}
\end{equation} 
Here, $\omega_{\text{max}}$ is a tunable threshold that sets the maximum bandwidth still considered "flat". Bands with energy spans exceeding this threshold automatically receive zero scores. To accommodate diverse materials and balance physical selectivity with data coverage, we consider multiple values of $\omega_{\text{max}}$ (100, 300, and 500 meV) in our experiments. The cosine transformation ensures that the score is maximized ($S_{\text{bandwidth}} = 1$) for a perfectly flat band and decreases smoothly to zero as the bandwidth approaches $\omega_{\text{max}}$. This non-linear decay emphasizes sensitivity to small variations in bandwidth at low values, consistent with the physical principle that strong correlation effects emerge when electronic kinetic energy is much smaller than the interaction energy. Bands with large dispersion rapidly lose relevance for correlated phenomena, and are assigned a low or zero score.

The DOS component, $S_{\text{DOS}}$, quantifies the concentration and peak prominence of electronic states within the representative band's energy range. We measure how concentrated the DOS is around the band's midpoint energy. The average DOS score of the local window and the broader surrounding DOS average within a fixed reference range of [-5 eV, 5 eV] \cite{regnault2022catalogue} are computed as follows:

\begin{equation}
\text{DOS}_\text{avg} = \frac{1}{N} \sum_{i=1}^{N} \text{DOS}(E_i), \quad \text{DOS}_\text{all} = \frac{1}{M} \sum_{j=1}^{M} \text{DOS}(E_j)
\end{equation}

with $E_i$ sampled in the local window, $E_i \in [E_{\text{mid}} - \frac{1}{2} \omega_{\text{max}}, E_{\text{mid}} + \frac{1}{2} \omega_{\text{max}}]$, and $E_j$ in a broad reference window of $[-5, 5]$ eV, $E_j \in [-5 eV, 5 eV]$. The final score $S_{\text{DOS}}$ is then calculated as follows:

\begin{equation}
S_\text{DOS} = \frac{\text{peak}_\text{contrast}}{1+\text{peak}_\text{contrast}},  \quad \text{where}  \quad \text{peak}_\text{contrast} = \frac{\text{DOS}_\text{avg}}{\text{DOS}_\text{all}}
\end{equation}

This formulation constrains $S_{\text{DOS}}$ to the range [0, 1], yielding a score of 0 when the local DOS is indistinguishable from the background average, and approaching 1 as a sharp peak emerges near the Fermi level. This provides a simple yet physically grounded indicator of localized electronic states that may enhance interaction-driven phenomena in flat-band systems.

The total flatness score combines the above components through a sigmoidal weighting scheme:
\begin{equation}
\begin{aligned}
S_{\text{total}} = 
\left\{
\begin{array}{ll}
0, & \text{if } \Delta E > \omega_{\text{max}}  \\
\text{sigmoid}\big(\lambda \cdot (S_\text{bandwidth}+S_\text{DOS})\big) \cdot & \\
\end{array}
\right. \\
\quad \text{sigmoid}\big(\beta \cdot (S_\text{bandwidth} \cdot S_\text{DOS})), \quad \text{otherwise}
\end{aligned}
\end{equation}

The first term rewards additive contributions, while the second accentuates cases where both dispersion suppression and DOS concentration co-occur. Parameters $\lambda$ and $\beta$ are tuned via Bayesian Optimization \cite{Nogueira2014}, using a target function shaped by HDBSCAN clustering \cite{mcinnes2017hdbscan}. 
The optimization objective linearly combines three terms:
(1) the mean $S_{\text{total}}$ within the high-density cluster (or 0 if absent), promoting high scores in structurally relevant regions;
(2) the mean $S_{\text{total}}$ in low-score regions (where either input falls below the 10th percentile), penalizing undesired elevation;
(3) the fraction of samples with $S_{\text{total}} > 0.95$, to prevent score saturation.
This formulation balances local fidelity with global regularization, encouraging physically meaningful score distributions. The final score is normalized to [0,1] and used as a regression target. Additional details of the flatness scoring formulation are provided in Supplementary Section A.

\subsection{Details of the deep Learning Model}

\subsubsection{Data Preprocessing}

To construct a multi-modal representation of 2D materials, we process data from 2DMatPedia and C2DB into graph and text modalities. Each entry provides crystal structure graph (the atomic species, lattice vectors, atomic positions) and textural property features exploration.

To represent the atomic structure of 2D materials, we convert each crystal into a graph-based representation following the ALIGNN framework. Each structure is expanded along the $c$-axis to suppress interlayer interactions under periodic boundary conditions:

\begin{equation}
    \mathbf{L}_c' = c_{\text{multi}} \cdot \mathbf{L}_c,
\end{equation}

where \( \mathbf{L}_c \) is the original lattice vector along the \( c \)-axis, and \( c_{\text{multi}} \) is a scaling factor that ensures a sufficiently large separation between periodic images. The transformed structure is then used to compute interatomic distances and neighboring relationships.

From the modified geometry, an undirected atomistic graph $G = (V, E)$ is constructed, where nodes correspond to atoms with CGCNN-derived elemental features, and edges represent interatomic bonds parameterized by interatomic distances. The $k$ nearest neighbors are identified adaptively based on in-plane lattice constants.
To encode local geometric environments beyond pairwise distances, we construct a line graph $L(G)$, where each node corresponds to a bond in $G$ and each edge encodes the angle between bond triplets:

\begin{equation}
    z_{(i,j)} = \cos(\theta_{(i,j,k)}),
\end{equation}

where \( \theta_{(i,j,k)} \) represents the angle between three connected atoms. This dual-graph formulation captures both radial and angular dependencies, allowing the model to learn from local geometric constraints crucial for electronic behavior in low-dimensional materials.

In addition to structural graphs, we incorporate tabulated crystal attributes, summarized in Table \ref{textlist}, to provide complementary information. These include symmetry descriptors, lattice parameters, atomic species, and formula representations. Each property is converted into a structured natural language sentence using a fixed template, forming a sequence of material-specific descriptors. The resulting text is tokenized and encoded using a pre-trained RoBERTa model\cite{liu2019roberta}, generating semantic embeddings for multi-modal integration.

\begin{table}
    \centering
    \label{textlist}
    \caption{Selected feature list. Each feature is accompanied by a specific description explaining its physical significance and contribution to material characterization.}
    \begin{tabular}{p{0.25\linewidth}p{0.7\linewidth}}
        \hline
        \textbf{Feature} & \textbf{Description} \\
        \hline
        Formula (Anonymous) & Generic chemical formula that removes specific element identifiers \\
        Formula (Pretty) & Human-readable chemical formula of the material \\
        Space Group & Space group indicating symmetry of the crystalline struct.~\\
        Point Group & Point group indicating symmetry around a point \\
        Crystal System & Geometric classification of the crystalline struct.~\\
        Lattice Matrix & Matrix representation of the lattice\\
        Lattice Parameters & Lattice constants and angles \\
        Atomic Species & Species of atoms in a unit cell \\
        Fractional Coordinates & The positions of atoms in the unit cell in fractional coordinates\\
        Lattice Volume & Volume of the unit cell based on lattice parameters \\
        \hline
    \end{tabular}
\end{table}

\subsubsection{Model Architecture}
We propose a deep learning Model that integrates atomic structures, textual property descriptions, and band-related features for predicting the flatness of electronic bands in two-dimensional materials. The model combines three primary components: a structure-aware graph encoder, a semantic-aware text encoder, and a bilinear interaction module to effectively fuse information from different modalities.

Our graph encoder is inspired by the ALIGNN framework \cite{choudhary2021atomistic} with a customized modular architecture captures both bond-level and angular interactions through coupled graph and line graph representations. 
In the first stage, Line-GNN layers update angle features $z$ and bond features $y$ by propagating information on the line graph $L(G)$, where each node represents a bond in the original atomic graph $G$. These updates incorporate angular dependencies among neighboring bonds. Simultaneously, atomic features $x$ are updated through their interactions with bonds:

\begin{equation}
\begin{aligned}
x' = x + \mathcal{F}{\text{node}}(x, y), 
\quad
y' = y + \mathcal{F}{\text{edge}}(y, z),
\end{aligned}
\end{equation}

where $x$, $y$, and $z$ denote atomic, bond, and angle features, respectively. This triplet-aware mechanism allows the model to encode higher-order geometric correlations essential to 2D material behavior. In the second stage, Gate GNN layers further refine node and bond embeddings via edge-gated convolutions. Bond features are updated using triplet-based angular information:

\begin{equation}
y_{ij}^{(l+1)} = y_{ij}^{(l)} + \mathrm{MLP} \left( \sum_{k \in \mathcal{N}(ij)} \sigma_{ijk} \cdot z_{ijk} \right)
\end{equation}

Updated bond embeddings guide the message passing among atoms, producing refined atomic features $x^{(l+1)}$. Finally, the graph-level representation is obtained by average pooling over atomic nodes:

\begin{equation}
h_{out} = \frac{1}{|V|} \sum_{i \in V} x_i^{\text{(final)}}
\end{equation}

This embedding serves as the structural input to the multimodal prediction module.

For the property representations, the textual input $T_\text{info}$ is first tokenized and passed through a pretrained RoBERTa model. The final embedding is extracted from the [CLS] token, representing the overall material description. This embedding is further projected through a linear layer with $\tanh$ activation and an MLP:

\begin{equation}
z_T = \text{MLP}(tanh(W_{lin}\cdot \text{RoBERTa}([CLS]))).
\end{equation}

This yields a compact semantic representation $z_T \in \mathbb{R}^d$ that encodes global text-derived properties.

To fuse the structural and semantic modalities, we first project both embeddings into a shared latent space and then apply a bilinear attention mechanism between the graph encoder output $h_\text{out}$ and the text embedding $z_T$. The joint representation is computed as:

\begin{equation}
z_F = h_G^\top W z_T,
\end{equation}

where $W \in \mathbb{R}^{d \times d \times d}$ is a learnable weight tensor. This interaction captures high-order correlations between atomic geometry and descriptive semantics. The fused vector $z_F$ is passed through a multi-layer perceptron to predict a scalar flatness score. This end-to-end architecture enables joint reasoning over atomic geometry and material semantics, providing accurate predictions for flat-band characteristics in two-dimensional materials.

\subsubsection{Training and Evaluation}

To robustly assess the performance of our model, we employed a 5-fold cross-validation strategy. In this setting, the dataset is partitioned into five subsets of approximately equal size. In each fold, 80\% of the data is used for training and 20\% for validation. We utilize the Adam optimizer with a scheduled learning rate reduction based on validation loss plateauing. For each epoch, model parameters are updated based on the mean squared error (MSE) between predicted and true flatness scores. Gradients are clipped to prevent explosion, and the model with the best validation \( R^2 \) score in each fold is saved for further analysis.

The model's performance is evaluated with a suite of regression metrics, including Root Mean Square Error ($RMSE$), Coefficient of Determination ($R^2$), and Mean Squared Error ($MSE$). The primary metric, $RMSE$, measures the average magnitude of prediction errors and is defined as:

\begin{equation}\label{eq_rmse}
	RMSE = \sqrt{\frac{1}{n} \sum_{i=1}^{n} (y_{ie} - y_{ip})^2},
\end{equation}

where \( y_{ip} \) and \( y_{ie} \) represent the predicted and experimental flatness scores, respectively, and \( n \) is the number of samples in the validation or test set. A smaller $RMSE$ indicates that the predicted values are close to the actual values. In addition, the $R^2$ metric assesses how well the model explains the variability of the response variable:

\begin{equation}\label{eq_r2}
	R^2 = 1 - \frac{\sum_{i=1}^{n} (y_{ie} - y_{ip})^2}{\sum_{i=1}^{n} (y_{ie} - \bar{y})^2},
\end{equation}

where \( \bar{y} \) is the mean of the observed flatness scores. An $R^2$ value closer to 1 indicates a better fit of the model to the data.

All predictions and evaluations are performed on unseen validation data in each fold, and the final performance is reported as the average across the five folds. This protocol ensures robust assessment of model performance and generalization.

\subsubsection{Experimental setting}
The deep learning model is implemented in Python 3.9, utilizing PyTorch 2.6.0 \cite{paszke2019pytorch} for neural network construction, DGL 1.1.0 \cite{wang2019dgl} for handling lattice graph structures, scikit-learn 1.6.1 \cite{pedregosa2011scikit} for data preprocessing, pymatgen 2024.8.9 \cite{ong2013pymatgen} for generating and analyzing 2D crystal structures, SciPy 1.13.1 \cite{virtanen2020scipy} for numerical optimization, and transformers 4.48.3 \cite{wolf2020transformers} for implementing the RoBERTa model to extract serialized crystal structure features.

For training, the batch size is set to be 8 and the Adam optimizer is used with a learning rate of 4e-1. We configured the model to use 5-fold cross-validation, with each fold running for a maximum of 100 epochs. 
For each iteration, the model's performance was monitored at every epoch on the validation set, tracking metrics $MSE$, $RMSE$ and $R^2$. The best epoch within each iteration, determined by the highest $R^2$, was selected. The final performance of the models was then calculated by averaging the metrics across all 5 iterations. This approach ensures that the evaluation captures the variability of the model's predictions, providing a reliable estimate of its performance. The configuration details and ablation analysis are provided in Supplementary Section B.

\subsection{Computational Validation}

To further prioritize candidates likely to exhibit kagome-induced flat bands, we applied a geometric filtering step that identifies kagome-like sublattice motifs based on coplanarity, bond length ratios, and local angular symmetry. Full algorithmic details are provided in Supplementary Section C.

We preformed DFT simulation using the Vienna Ab Initio Simulation Package (VASP) to calculate electronic properties including band structures and Kohn-Sham wavefunctions. We utilized the projector augmented wave (PAW) method and the Perdew–Burke–Ernzerhof (PBE) exchange-correlation functional, similar as in the Materials Project \cite{Perdew1996-dk, jain2011high}. To prepare input files and plot band structures after DFT simulation, Pymatgen codebase is used \cite{jain2011high, ong2013python}. To compute irreducible representations, we employed the protocol described in \cite{gao2021irvsp} using "irvsp" to calculate irreps from wavefunctions and "phonopy" \cite{phonopy-phono3py-JPCM, phonopy-phono3py-JPSJ} and "pos2aBR" \cite{gao2022unconventional, nie2021application} to prepare standard POSCAR files. The Bilbao Crystallographic Server \cite{aroyo2011crystallography, aroyo2006bilbao1, aroyo2006bilbao2} is used to compare band representations with irreps of EBRs of layer groups \cite{petralanda2024two}. To draw band structures from simulation with spin-orbit coupling, "pyprocar" codebase \cite{herath2020pyprocar, lang2024expanding} is used.

\section*{Date availability}
The dataset used for training the deep learning models, along with the set of high-scoring candidate materials identified by our framework, is available on GitHub at https://github.com/Xiangwen-Wang/Struct2Flat.

\section*{Code availability}
All code used for flatness score computation, deep learning model training, and candidate screening is available at https://github.com/Xiangwen-Wang/Struct2Flat.

\section*{Acknowledgments}
We gratefully acknowledge financial support from UK Research and Innovation Grant [EP/X017575/1], European Research Council (ERC) under the European Union's Horizon 2020 research and innovation program (Grant Agreement No. 865590), the Royal Society University Research Fellowship URF$\backslash$R1$\backslash$221096 and the Research Council UK [BB/X003736/1].

\section*{Author information}
Authors and Affiliations

Department of Physics and Astronomy, University of Manchester, Manchester, UK

Xiangwen Wang, Yihao Wei, Anupam Bhattacharya1, Qian Yang and Artem Mishchenko

Contributions

A.M. and Q.Y. initiated and supervised the project. X.W. designed the methodology for flatness scoring and developed the multimodal deep learning model. X.W. implemented the codebase and conducted the primary experiments. X.W. and Y.W. curated the dataset, including data extraction and preprocessing. Y.W. performed the first-principles validation. X.W. and Y.W. prepared the figures and visualizations, and drafted the manuscript. A.M. and Q.Y. contributed to the final conceptual framing and correlation analysis. All authors discussed the results and commented on the paper.

Corresponding authors

Correspondence to Xiangwen Wang, Qian Yang or Artem Mishchenko.

\section*{Ethics declarations}
The authors declare no competing interests.


\bibliography{sn-bibliography}

\end{document}